\documentclass[aip, apl, reprint, showkeys, showpacs]{revtex4-1}
\usepackage[T1]{fontenc}
\usepackage[latin9]{inputenc}
\usepackage{verbatim}
\usepackage{graphicx}
\usepackage[english]{babel}
\makeatletter
\DeclareRobustCommand*\textsubscript[1]{%
  \@textsubscript{\selectfont#1}}
\def\@textsubscript#1{%
  {\m@th\ensuremath{_{\mbox{\fontsize\sf@size\z@#1}}}}}
\makeatother

\begin{document}

\title{Strain relaxation by dislocation glide in ZnO/ZnMgO core-shell nanowires}

\author{G. Perillat-Merceroz}
\email{guillaume.perillat@epfl.ch}
\affiliation{CEA, LETI, Minatec Campus, Grenoble, 38054, France}
\affiliation{CEA INAC/UJF-Grenoble1 UMR-E, SP2M, LEMMA, Minatec
Campus, Grenoble, 38054, France}
\author{R. Thierry}
\affiliation{CEA, LETI, Minatec Campus, Grenoble, 38054, France}
\author{P.-H. Jouneau}
\affiliation{CEA INAC/UJF-Grenoble1 UMR-E, SP2M, LEMMA, Minatec
Campus, Grenoble, 38054, France}
\author{P. Ferret}
\affiliation{CEA, LETI, Minatec Campus, Grenoble, 38054, France}
\author{G. Feuillet}
\affiliation{CEA, LETI, Minatec Campus, Grenoble, 38054, France}

\begin{abstract}
Plastic relaxation of the misfit stress in core-shell semi-conducting
nanowires can lead to structural defects, detrimental to
applications. Core-shell Zn\textsubscript{0.7}Mg\textsubscript{0.3}O/ZnO quantum well heterostructures were deposited on ZnO nanowires. Strain along the \textbf{a} and \textbf{c} axes of the wurtzite structure is relaxed through the glide of dislocation half-loops from the free
surfaces, within pyramidal and prismatic planes. Some half-loops are
closed up in the barriers to accommodate
the misfit at two consecutive interfaces of the quantum well stack.
Dislocations are also observed within the nanowire core: contrary to two-dimensional structures, both the core and the
shell can be plastically relaxed.
\end{abstract}

\maketitle

Due to their particular properties, semi-conducting nanowires are
widely studied for nanoelectronics\cite{huang_logic_2001} and optoelectronics.\cite{huang_nanowires_2005}
The core-shell configuration is especially interesting for various
applications such as transistors made of silicon and germanium,\cite{lauhon_epitaxial_2002}
solar cells made of a ZnO core and a ZnSe shell,\cite{wang_direct_2008}
or light-emitting diodes with GaN/InGaN core-shell quantum wells.\cite{bavencove_light_2011}
Although the core-shell structures described in most of the published
works are coherently strained,\cite{kavanagh_misfit_2010} the stress
due to the lattice mismatch between the core and the shell materials
may plastically relax if thicknesses and/or alloy concentrations exceed
some critical values. This leads to the formation of misfit dislocations
which are detrimental to device efficiencies. For example, it was
demonstrated that internal quantum efficiencies of ZnO core-shell
quantum wells could vary from 1 to 54\% depending on the presence
or not of dislocations,\cite{thierry_core-shell_2012} and that InAs/GaAs core-shell nanowires containing dislocations had a reduced electron field-effect mobility compared to bare InAs nanowires.\cite{kavanagh_transport_2011} It is therefore
important to understand, and control, the strain relaxation mechanisms.

Strain relaxation in cubic two-dimensional (2D) heterostructures has
been considered decades ago: it usually occurs by introducing misfit
dislocations at the heterostructure interface, either by bending of
pre-existing dislocations, or through the glide of dislocation half-loops
from the free surface, generally on the densest atomic planes.\cite{matthews_defects_1974,matthews_defects_1975}
Similar mechanisms are at play when strain relaxation occurs through
dislocations in hexagonal compact 2D heterostructures.
For \emph{c} oriented growth, glide occurs in pyramidal planes, because
the dense basal and prismatic planes are not active glide systems.\cite{srinivasan_slip_2003,floro_misfit_2004}
For \emph{m} plane growth, glide can occur on prismatic planes,\cite{yoshida_evidence_2011}
and for semipolar plane growth, glide can occur on both basal and
prismatic planes.\cite{wu_observation_2011} Concerning relaxation
in nanowire core-shell structures, theoretical studies have been published.\cite{ovidko_misfit_2004,raychaudhuri_critical_2006,trammell_equilibrium_2008}
Strain energies have been calculated for cubic\cite{trammell_equilibrium_2008}
and wurtzite\cite{raychaudhuri_critical_2006} structures. For wurtzite
materials, the strain energy was compared to the dislocation energy
in order to determine the critical thickness for plastic relaxation.
The role of the shell thickness and composition was shown, similarly
to the 2D case, but the core thickness has also an influence because
both the core and the shell are stressed and store a part of the strain
energy.\cite{raychaudhuri_critical_2006} Numerical evaluation of
the critical thicknesses was made for nitride materials. In these models, edge misfit dislocations were assumed to be formed by glide and climb.\cite{ovidko_misfit_2004}
Concerning experimental observations, edge misfit dislocations were observed at the interface of wurtzite InAs/GaAs core-shell nanowires.\cite{popovitz-biro_inas/gaas_2011,kavanagh_faster_2012} However, there are no indications of dislocations threading from the InAs/GaAs interface to the surface, which made more difficult discussing the glide systems. Concerning ZnO/ZnMgO core-shell nanowires we have shown that structures with Zn\textsubscript{0.85}Mg\textsubscript{0.15}O
barriers on ZnO were elastically strained, contrary to structures
with Zn\textsubscript{0.7}Mg\textsubscript{0.3}O barriers, which
were plastically relaxed because of a twice higher lattice mismatch.\cite{thierry_core-shell_2012}
The type of the formed dislocations was not addressed for ZnO/ZnMgO, and the glide systems
were addressed neither for InAs/GaAs nor for ZnO/ZnMgO.

In this work, we first analyze the possible glide systems for dislocations
in core-shell nanowires with the wurtzite structure. Then, the dislocations
caused by plastic relaxation are experimentally observed in a system
composed of a ZnO core, and of a heterostructure shell consisting
in four Zn\textsubscript{0.7}Mg\textsubscript{0.3}O barriers and
three ZnO quantum wells. It is shown that the misfit strain is relaxed
through the glide of dislocation half-loops from the surface in prismatic
and pyramidal planes. Some loops are closed because of the presence
of alternated ZnMgO and ZnO shell layers. Finally, a few dislocations
are also observed in the ZnO core, showing that both the shell and
the core relax to accommodate the strain.

ZnO nanowires were grown on $\left(0001\right)$ sapphire by metal
organic vapor phase epitaxy: details about the growth can be found
in a previous publication.\cite{rosina_morphology_2009} The nanowires
grew vertically along the \textbf{+c} direction, without any structural
defects such as stacking faults or dislocations.\cite{perillat-merceroz_mocvd_2010,perillat-merceroz_compared_2012}
ZnO core-shell multi-quantum wells were subsequently
grown radially on the ZnO nanowires. The heterostructure consists
in three 4~nm-thick ZnO quantum wells within 10~nm-thick Zn\textsubscript{0.7}Mg\textsubscript{0.3}O
barriers.\cite{thierry_core-shell_2012} The lattice mismatch between Zn\textsubscript{0.7}Mg\textsubscript{0.3}O
and ZnO is +0.6\% along \textbf{a}, -0.7\% along \textbf{c}, and Zn\textsubscript{1-x}Mg\textsubscript{x}O was shown to be of wurtzite structure up to x=0.33.\cite{ohtomo_mgzno_1998} In our case with x=0.3, it is thus expected that ZnMgO keep the wurtzite structure. Electron diffraction patterns showed only one set of spots corresponding to the wurtzite phase for the ZnO/ZnMgO heterostructure, because the misfit between ZnO and ZnMgO is too small to distinguish separated spots for the two materials. Transmission electron microscopy (TEM) images were taken on a FEI-Tecnai
microscope operated at 200~kV. TEM samples were prepared by the cleaved edge method for cross sections, and by focused ion
beam for plan views. Details about the growth conditions and TEM sample preparation can be found in Ref.\cite{thierry_core-shell_2012}. The $\left\langle 0001\right\rangle $,
$\frac{1}{3}\left\langle 11\bar{2}0\right\rangle $, and $\frac{1}{3}\left\langle 1\bar{1}00\right\rangle $
directions of the hexagonal compact structure are noted \textbf{c},
\textbf{a}, and \textbf{p} respectively. The $\left\{ 0001\right\} $
basal plane is refered as the \emph{c} plane, while the prismatic
planes $\left\{ 11\bar{2}0\right\} $ and $\left\{ 1\bar{1}00\right\} $
are the \emph{a} and \emph{m} planes respectively. In the hexagonal
compact structure, the more common dislocations are either perfect
ones, with Burgers vector\textbf{ b} equal to \textbf{a}, \textbf{c},
or \textbf{a}+\textbf{c}, and partial ones with \textbf{b} equal to
\textbf{p}, 2\textbf{p}, \textbf{c}/2, \textbf{c}+2\textbf{p}, \textbf{c}/2+\textbf{p},
or \textbf{c}+\textbf{p}.\cite{hirth_theory_1982}

\begin{figure}
\begin{centering}
\includegraphics[width=8cm]{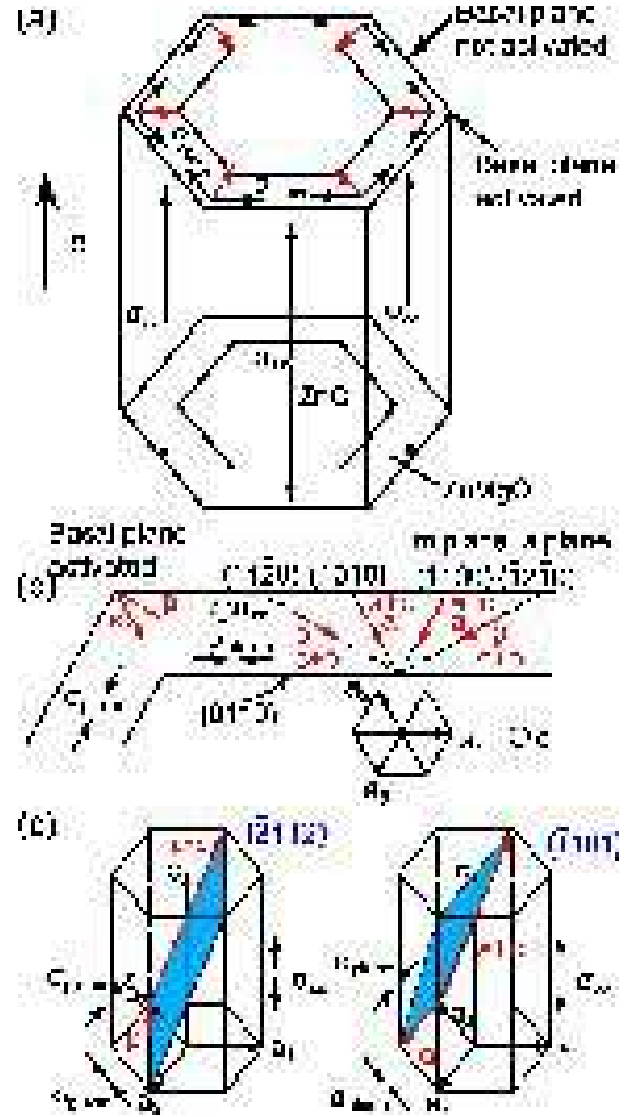}
\par\end{centering}
\caption{(a) Stress in a wurtzite core-shell nanowire: the lattice misfit between
the ZnO core and the ZnMgO shell lead to a vertical stress $\sigma_{zz}$
and a planar stress $\sigma_{plane}$ (black arrows). At the corners
between two \emph{m} facets of the nanowire, there is a component
of the planar stress directed towards the core (red arrows). (b) and
(c): glide systems with possible glide directions indicated by
red arrows. (b) The activated prismatic glide planes are represented
by dashed lines, and the basal plane is activated only at the corner.
(c) Some activated pyramidal glide planes are colored in blue.}
\raggedright{}\label{glide_systems}
\end{figure}

Before describing our observations of dislocations due to plastic
relaxation, let us examine what the activated glide systems could be
for dislocations in wurtzite core-shell nanowires with \emph{m}-plane
facets. The stress state of a wurtzite nanowire consisting of a ZnO
core and a ZnMgO barrier is presented in figure \ref{glide_systems}
(a). The actual sample studied here is more complicated, with a stack
of four ZnO quantum wells and three ZnMgO barriers grown on a ZnO
core. However, in a first approach, this stack can be approximated
by a unique ZnMgO barrier with an average Mg content lower than the
four ZnMgO barriers. ZnMgO has a \emph{c} parameter smaller than ZnO,
and a larger \emph{a} parameter.\cite{ohtomo_mgzno_1998} This lattice
mismatch induces a tensile stress in the ZnMgO shell along the \textbf{c}
direction, and a compressive stress in the \textbf{a} direction, for
each \emph{m} facet of the nanowire. Consequently, at the corner between
two facets, there is a component of the stress directed towards the
center of the nanowire, which is not present in the case of a 2D layer
grown on a \emph{m} face. Figure \ref{glide_systems} (b) represents
the possible prismatic glide planes by dashed lines in a core-shell nanowire, with
the activated glide directions indicated by red arrows. \emph{a} and
\emph{m} planes can be activated by the misfit stress, with glide
directions being respectively \textbf{p} or \textbf{c}+\textbf{p},
and \textbf{a} or \textbf{a}+\textbf{c}. Burgers vector can be respectively
\textbf{c},\textbf{ p} or \textbf{c}+\textbf{p}, and \textbf{c}, \textbf{a}
or \textbf{c}+\textbf{a}. Figure \ref{glide_systems} (c) shows possible
pyramidal glide planes and glide directions. Dislocations can glide
for example in $\left\{ \bar{2}112\right\} $ planes along \textbf{p}
or \textbf{a}+\textbf{c} with \textbf{b}=\textbf{p}, \textbf{a}+\textbf{c},
\textbf{c}+2\textbf{p}, or \textbf{c}/2+\textbf{p}, and in $\left\{ \bar{1}101\right\} $
planes along \textbf{a} or \textbf{a}+\textbf{c} with \textbf{b}=\textbf{a}
or \textbf{a}+\textbf{c}. Higher index pyramidal planes could also
be envisaged, but they would be energetically less favored. Because
the stress is directed inwards at the corners of the nanowire, the
projected stress on the prismatic and pyramidal planes is increased
at these corners. Moreover, glide in the basal plane directions \textbf{p} and \textbf{a} are activated
at these corners, with a Burgers vector \textbf{p} or \textbf{a}. Besides, it was observed that, for GaN/InGaN 2D heterostructures
grown on semipolar planes, glide of dislocations with \textbf{b}=\textbf{a}
can occur in the basal planes.\cite{wu_misfit_2011} Consequently,
the misfit dislocations with \textbf{b}=\textbf{c}/2 in a basal plane observed in GaAs/InAs core-shell nanowires\cite{popovitz-biro_inas/gaas_2011,kavanagh_faster_2012} cannot be formed by glide from the free surface. It was proposed that this type of dislocation could be formed by climb, by dislocation reaction, or by glide.\cite{ovidko_misfit_2004,kavanagh_faster_2012} Glide would be possible only at the very top of the nanowires, or if the shell growth occurred by isolated islands which would coalesce.

\begin{figure}
\begin{centering}
\includegraphics[width=8cm]{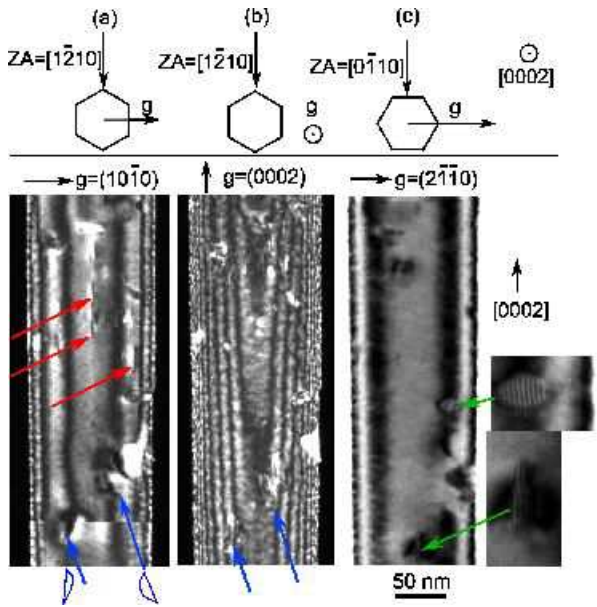}
\par\end{centering}
\caption{Weak-beam TEM images of a core-shell nanowire taken along the $\left[1\bar{2}10\right]$
zone axis (ZA), with (a) $g=\left(10\bar{1}0\right)$, and (b) $g=\left(0002\right)$.
(c) Two-beam TEM image taken along the $\left[0\bar{1}10\right]$
zone axis with $g=\left(2\bar{1}\bar{1}0\right)$. The red arrows indicate
dislocation loops in prismatic planes, the blue ones indicate
dislocation loops in pyramidal planes, and the green ones indicate moirés inside
dislocation loops on the $g=\left(2\bar{1}\bar{1}0\right)$ image, with zooms on these moirés.}
\label{cross_section}
\end{figure}

Figure \ref{cross_section} shows weak-beam TEM images of the same
nanowire taken along the $\left[1\bar{2}10\right]$ zone axis, with
(a) $g=\left(10\bar{1}0\right)$ and (b) $g=\left(0002\right)$, and
(c) a two-beam image taken along the $\left[0\bar{1}10\right]$ zone
axis with $g=\left(2\bar{1}\bar{1}0\right)$. Different types of dislocations
are visible.

First, dislocation half-loops in pyramidal planes are indicated by
blue arrows on the $g=\left(10\bar{1}0\right)$ and $g=\left(0002\right)$
images. Their shape is outlined in blue for more clarity. One must
keep in mind that TEM images are 2D projections of the whole nanowire:
that is the reason why some half-loops do not continue until the apparent
sides of the nanowire. For $g=\left(2\bar{1}\bar{1}0\right)$, vertical
fringes may be visible in the pyramidal plane dislocations (some are
indicated by green arrows). This fringes are a moiré caused by a stacking
fault inside the dislocation half-loop. However, stacking faults caused
by the glide of common partial dislocations in the wurtzite structure
(\textbf{p}, 2\textbf{p}, \textbf{c}/2, \textbf{c}+2\textbf{p}, \textbf{c}/2+\textbf{p},
\textbf{c}+\textbf{p}) are extinguished with $g=\left(2\bar{1}\bar{1}0\right)$,
according to the criterion which states that stacking faults are not
visible if $2\pi\,\mathbf{g}\cdot\mathbf{R}$ is a multiple of $2\pi$
(with $\mathbf{R}$ the displacement vector of the stacking
fault). Furthermore, prismatic stacking faults with $\mathbf{R}=\mathbf{c}/2+\frac{3}{2}\mathbf{p}$
have been observed in the wurtzite structure, and give a contrast
for $g=\left(2\bar{1}\bar{1}0\right)$.\cite{drum_intersecting_1965} These prismatic stacking faults
intersected basal ones with $\mathbf{R}=\mathbf{c}/2+\mathbf{p}$,
and thus stair-rod dislocations with \textbf{b}=\textbf{p}/2 were
present at the intersections. We speculate that the dislocations observed
here may have a non classical Burgers vector component in the basal plane such as \textbf{b}=\textbf{p}/2,
which gives a stacking fault contrast with $g=\left(2\bar{1}\bar{1}0\right)$.
Because these dislocations are visible for $g=\left(10\bar{1}0\right)$
and $g=\left(0002\right)$, their Burgers vector \textbf{b} has not only a component in the basal plane, but also a component along \textbf{c}. Consequently, they release the misfit strain both along \textbf{a}
and along \textbf{c}. It is worth noting that the major part of the dislocations is threading to the surface, whereas the portion of the dislocations at the interface (the so-called "misfit dislocations") is relatively shorter. This is in contrast with the observations in InAs/GaAs nanowires where  supplementary half-planes due to large misfit dislocations were visible on high resolution TEM images.\cite{popovitz-biro_inas/gaas_2011,kavanagh_faster_2012} This difference could be due to the different values of misfit: 0.6\% for ZnO/Zn\textsubscript{0.7}Mg\textsubscript{0.3}O, and 6.8\% for InAs/GaAs, along \textbf{a}. Moreover, it was shown that some inclined threading dislocations do play a role in the strain relaxation.\cite{cantu_role_2005}


Second, dislocation half-loops in $\left(10\bar{1}0\right)$ planes which give a contrast along a vertical line on the image with $g=\left(10\bar{1}0\right)$ are indicated by red arrows. Considering perfect dislocations, the three possibilities for the Burgers vector for a dislocation gliding in the $\left(10\bar{1}0\right)$ plane are $\mathbf{b}=\left[1\bar{2}10\right]=\mathbf{a}$, $\mathbf{b}=\mathbf{c}$, and $\mathbf{b}=\frac{1}{3}\left[1\bar{2}13\right]=\mathbf{a}+\mathbf{c}$. Because the dislocations are not visible for $g=\left(0002\right)$, they should be $\mathbf{b}=\left[1\bar{2}10\right]$, according to the simplified extinction criterion which states that a dislocation is not visible when $\mathbf{g}\cdot\mathbf{b}=0$. However, a dislocation with $\mathbf{b}=\left[1\bar{2}10\right]$ should neither be visible with $g=\left(10\bar{1}0\right)$. Besides, they are not visible on the bright field image taken with $g=\left(10\bar{1}0\right)$ (not shown
here). The contrast on the weak-beam $g=\left(10\bar{1}0\right)$ image could be a residual contrast because the term $\mathbf{g}\cdot\mathbf{b}\wedge\mathbf{u}$
is not equal to zero for $\mathbf{b}=\left[1\bar{2}10\right]$ and \textbf{u} along \textbf{c}. These dislocations can release misfit strain along the \textbf{a} direction. Concerning 2D heterostructures, it was observed for GaN/InGaN
that relaxation could occur through the glide of dislocations with
\textbf{b}=\textbf{a} on prismatic planes, for growth on \emph{m}
plane\cite{yoshida_evidence_2011} and semipolar plane\cite{wu_misfit_2011}. Moreover, \textbf{b}=\textbf{a} misfit dislocations were observed in InAs/GaAs core-shell nanowires.\cite{popovitz-biro_inas/gaas_2011}

To summarize, dislocations are observed to glide in prismatic and pyramidal
planes, but not in basal planes. Consequently, the misfit along \textbf{c
}is not released in our case by dislocations lying in basal planes, as proposed in theoretical models,\cite{ovidko_misfit_2004,raychaudhuri_critical_2006} and as observed for InAs/GaAs core-shell nanowires.\cite{popovitz-biro_inas/gaas_2011,kavanagh_faster_2012} Instead, it is released by loops lying in pyramidal planes with a Burgers vector
with two components, one along \textbf{c}, and one within the \emph{c}
plane. Misfit along \textbf{a} is released by these loops in pyramidal planes, and by loops in prismatic planes.

\begin{figure}
\begin{centering}
\includegraphics[width=8cm]{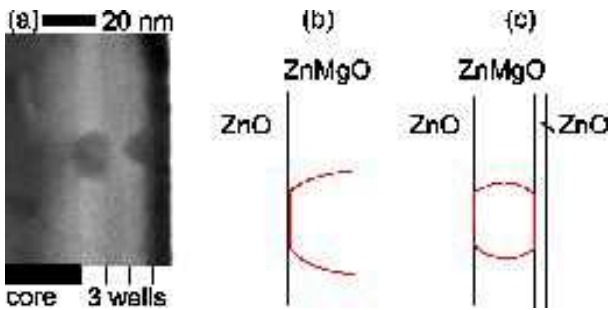}
\par\end{centering}
\caption{(a) Two-beam TEM image with $\mathbf{g}=\left(10\bar{1}0\right)$
of a closed dislocation loop within a ZnMgO barrier, (b) schematics
of a half-dislocation loop formation in ZnMgO, and (c) schematics
of the closed dislocation loop between the ZnO core and the ZnO quantum
well.}
\raggedright{}\label{closed_loops}
\end{figure}

Figure \ref{closed_loops}(a) shows another TEM image of a core-shell
nanowire viewed in cross-section: the ZnO core is visible on the left,
and then the four ZnMgO barriers and the three ZnO quantum wells.
A closed dislocation loop is observed between the ZnO core and the
first ZnO quantum wells. Its formation mechanism can be explained
as follows. First, a dislocation half-loop is formed in a ZnMgO barrier
to accommodate the misfit with the ZnO core [Fig. \ref{closed_loops}(b)]. Then, because of the
subsequent growth of a ZnO quantum well on the ZnMgO barrier, the
dislocation half-loop closes up in order to accommodate the misfit
between the ZnMgO barrier and the ZnO well: the same dislocation loop
accommodates the misfit at the two opposite ZnO/ZnMgO interfaces [Fig. \ref{closed_loops}(c)].
A similar explanation was originally proposed for a GaAs/GaAsP 2D
multiple-stack.\cite{matthews_defects_1974}

\begin{figure}
\begin{centering}
\includegraphics[width=6.5cm]{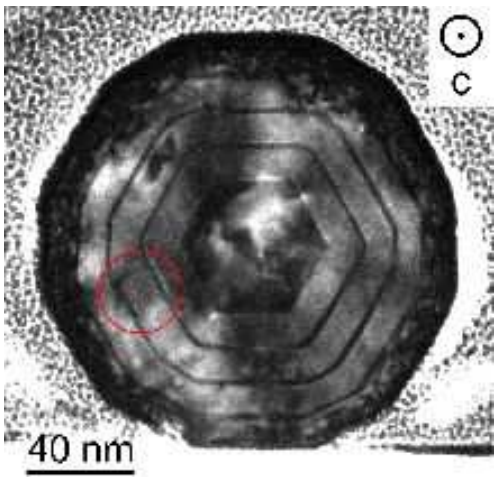}
\par\end{centering}
\caption{Plane view TEM image in the $\left[0001\right]$ zone axis with an
aperture to select only the direct beam. A closed dislocation loop
with a moiré is visible between two ZnO quantum wells (circled in
red), and a dislocation is visible in the ZnO core.}
\label{Moire}
\end{figure}

Figure \ref{Moire} shows a plan view image of a nanowire viewed
along the \textbf{$\left[0001\right]$} zone axis, with an objective
aperture on the direct beam. A pyramidal dislocation loop with moiré
fringes, circled in red, is visible between two ZnO quantum wells:
it is another example of a closed dislocation loop. Moreover, a dislocation
is visible in the nanowire core. Trammell \emph{et al.} showed that
the strain energy was mainly stored in the shell for thin shells relative
to the core, whereas this energy was mainly stored in the core for
relatively thicker shells.\cite{trammell_equilibrium_2008} Consequently,
there would be two steps for the plastic relaxation in core-shell
structures. First, when the shell is thin compared to the ZnO core,
plastic relaxation occurs mainly in the shell. Second, when the shell
becomes thicker, the ZnO core itself may relax to accommodate the
misfit strain through the generation of dislocations within it. However,
the source of these dislocations is not obvious because the ZnO core
is free of defects before the growth of the shell. It is worth noting
that according to Raychaudhuri \emph{et al.}, with a thin
enough core and a low enough barrier composition, the barrier can
have an infinite thickness without being plastically relaxed.\cite{raychaudhuri_critical_2006} Our
observations in Ref.\cite{thierry_core-shell_2012} seems to confirm
this point: for 75~nm thick barriers with 15\% Mg concentration on
a 60~nm core, no plastic relaxation was observed.

To conclude, we have identified plastic relaxation mechanisms in ZnO/ZnMgO
core-shell nanowires. It was shown, through a contrast analysis in
TEM, that strain relaxation along the\textbf{ c} and \textbf{a} directions
of the facets of the wurtzite core-shell nanowires occurred through
the glide of dislocation half-loops from the free surface to the interface.
Glide occurs in pyramidal planes, with the Burgers vectors
of the dislocations having a component both along \textbf{c} and in
the \emph{c} plane, and in prismatic planes, with \textbf{b}=\textbf{a}. Some of these half-loops may close up in order
to accommodate the misfit at two consecutive opposite interfaces (ZnO/ZnMgO
and ZnMgO/ZnO) of the nanowire heterostructure. The stress state in these one-dimensional structures is different from 2D layers, because of
the finite dimensions of the \emph{m} and \emph{a} facets. Actually,
the presence of corners between these facets induces inhomogeneous
strain fields in the shell. Another
difference is the presence of dislocations within the nanowire core
itself, although the formation mechanism of these dislocations remains
unclear. These observations will help design optimized core-shell
nanowires heterostructures, by an adequate choice of core and layer
thicknesses and composition in order to avoid the formation of detrimental
defects such as dislocations.

\begin{acknowledgments}
The authors acknowledge Eric Gautier for help during the FIB preparation
of TEM samples, and the French national research agency (ANR)
for funding through the Carnot program (2006/2010).
\end{acknowledgments}

\bibliography{biblio}

\end{document}